\documentclass[aps,prl,twocolumn,groupedaddress,superscriptaddress,amsfonts,amssymb,amsmath,citeautoscript,a4paper]{revtex4-2}

\usepackage{orcidlink}
\usepackage{lipsum}
\usepackage[utf8]{inputenc}
\usepackage[english]{babel}
\usepackage{microtype}
\usepackage{txfonts}
\usepackage{txfontsb}
\usepackage{bbm}
\usepackage{xcolor}
\usepackage{graphicx}
\usepackage{float}
\usepackage{xspace}
\usepackage[skip=0pt, indent=15pt]{parskip}
\usepackage{titlesec}
\titleformat{\paragraph}[runin]{\normalfont\normalsize\bfseries}{}{0}{}[]
\titleformat{\section}[hang]{\normalfont\large\bfseries}{}{0}{}[]
\usepackage{subfiles}
\usepackage{enumerate}
\usepackage[inline]{enumitem}
\usepackage{multirow,rotating}

\usepackage{siunitx}
\DeclareSIUnit[number-unit-product=]\percent{\char`\%} % remove space before percentage "units"

\hypersetup{colorlinks,
            linkcolor={blue!75!black!80!yellow},
            citecolor={blue!75!black!80!yellow},
            urlcolor={blue!75!black!80!yellow},
            pdfstartview=FitH}

\makeatletter
\renewcommand{\fnum@figure}{FIG.~\thefigure}
\makeatother

\usepackage{xifthen}
\usepackage{etoolbox}
\usepackage{soul}

\usepackage{mdframed}
\definecolor{blue-violet}{rgb}{0.54, 0.17, 0.89}

\newmdenv[topline=false, rightline=false, bottomline=false,%
  linewidth=1.25pt, innerrightmargin=0pt, leftmargin=-8pt,%
  innerleftmargin=7pt, skipabove=0pt, skipbelow=0pt,%
  linecolor=blue-violet, fontcolor=blue-violet]{mdleftbar}

\begin{document}

\title{Local control and lateral nanofocusing of hyperbolic phonon polaritons}

\author{Jacob~T.~Heiden\,\orcidlink{0000-0003-4505-6107}}
\affiliation{School of Electrical Engineering, Korea Advanced Institute of Science and Technology (KAIST), Daejeon 34141, Korea}

\author{{Haozhe~Tong},\orcidlink{0009-0008-9053-9312}}
\affiliation{Donostia International Physics Center (DIPC), Donostia-San Sebastián 20018, Spain}
\affiliation{Universidad del Pais Vasco/Euskal Herriko Unibertsitatea (UPV/EHU), Donostia-San Sebastián 20018, Spain}

\author{Yongjun~Lim}
\affiliation{Department of Biomicrosystem Technology, Korea University, Seoul 02841, Korea}

\author{Heerin~Noh}
\affiliation{KU-KIST Graduate School of Converging Science and Technology, Korea University, Seoul 02841, Korea}

\author{Pablo~Alonso-Gonz\'{a}lez\,\orcidlink{0000-0002-4597-9326}}
\address{Department of Physics, University of Oviedo, Oviedo 33006, Spain}
\address{Center of Research on Nanomaterials and Nanotechnology, CINN (CSIC-Universidad de Oviedo), El Entrego 33940, Spain}

\author{Alexey.~Y.~Nikitin\,\orcidlink{0000-0002-2327-0164}}
\address{Donostia International Physics Center (DIPC), Donostia-San Sebastián 20018, Spain}
\address{IKERBASQUE, Basque Foundation for Science, Bilbao 48013, Spain}

\author{Seungwoo~Lee\,\orcidlink{0000-0002-6659-3457}}
\affiliation{Department of Biomicrosystem Technology, Korea University, Seoul 02841, Korea}
\affiliation{KU-KIST Graduate School of Converging Science and Technology, Korea University, Seoul 02841, Korea}
\affiliation{Department of Integrative Energy Engineering (College of Engineering), Korea University, Seoul 02841, Korea}

\author{Sergey~G.~Menabde\,\orcidlink{0000-0001-9188-8719}}
\thanks{\href{mailto:menabde@kaist.ac.kr}{menabde@kaist.ac.kr}; \href{mailto:jang.minseok@kaist.ac.kr}{jang.minseok@kaist.ac.kr}}
\affiliation{School of Electrical Engineering, Korea Advanced Institute of Science and Technology (KAIST), Daejeon 34141, Korea}
\email{menabde@kaist.ac.kr}\email{jang.minseok@kaist.ac.kr}

\author{Min~Seok~Jang\,\orcidlink{0000-0002-5683-1925}}
\thanks{\href{mailto:menabde@kaist.ac.kr}{menabde@kaist.ac.kr}; \href{mailto:jang.minseok@kaist.ac.kr}{jang.minseok@kaist.ac.kr}}
\affiliation{School of Electrical Engineering, Korea Advanced Institute of Science and Technology (KAIST), Daejeon 34141, Korea}
\email{menabde@kaist.ac.kr}\email{jang.minseok@kaist.ac.kr}

\begin{abstract}
Phonon polaritons in van der Waals crystals enable exceptional light confinement and control over low-loss nanolight propagation. The polariton wavelength can be controlled by the crystal geometry, isotopic composition, or surrounding environment -- for which substrate engineering is particularly effective. However, existing approaches of substrate nanopatterning are binary and offer limited leverage. Here, we demonstrate local control over the wavelength of phonon polaritons in hexagonal boron nitride by employing a sinusoidally corrugated gold surface to smoothly vary the gap between the van der Waals crystal and metallic substrate. The nonuniform gap provides a continuous and nearly threefold local variation of the polariton wavelength across the structure, verified by near-field optical microscopy. Our platform further enables lateral nanofocusing by gradually compressing and decompressing the wavelength of propagating polaritons by a factor of around 2.5 achieved solely through substrate geometry, consistent with our local control experiments and theoretical calculations. Our results push the boundaries of substrate engineering and showcase a powerful method for precise and local tailoring of polaritonic modes.
\end{abstract}

\maketitle

%\section{Introduction}
The ability to tailor light at the nanoscale has become of great importance, being a cornerstone in modern photonics. To this end, two-dimensional (2D) van der Waals (vdW) crystals -- such as graphene and hexagonal boron nitride (hBN) -- have emerged as key materials, to a large extent due to the polaritons they support. Polaritons are hybrid light-matter modes that are born out of coupling between photons and material charges, thereby enabling nanoscale light confinement and control~\cite{Basov:2016, Low:2017, Basov:2021}. In particular, polar vdW crystals such as hBN support hyperbolic phonon polaritons (HPhPs) which dispersion in the infrared regime can be modified by the material geometry~\cite{Caldwell:2014}, crystal thickness~\cite{Dai:2014}, its isotopic composition~\cite{Giles:2018}, and surrounding environment~\cite{Dai:2019,Sternbach:2023}. Such versatile tunability enables engineering of HPhP propagation and wavefronts at deeply subwavelength scales. Combined with a strong electromagnetic field confinement, this unlocks novel applications in molecular sensing~\cite{Autore:2018,Bylinkin:2021}, nanoscale energy flow steering~\cite{Maia:2019}, nonlinear optics~\cite{Ginsberg:2023}, and sub-diffraction imaging~\cite{Dai:2015,Li:2015}.

Among the levers that tune polaritonic dispersion, modifying the surrounding environment has been shown to be uniquely effective. In particular, placing hBN in close proximity to a metallic substrate leads to the manifestation of hyperbolic image phonon polaritons~\cite{Ambrosio:2018,Lee:2020,Menabde:2022b}, arising from the screening of polaritonic fields by image charges in the metal. This screening significantly alters the dispersion, with image modes exhibiting about three times larger in-plane momentum, $q$, compared to HPhPs in a freestanding hBN~\cite{Menabde:2022a}. Beyond uniform environmental tuning, spatial variation of dielectric environment offers an additional degree of freedom for manipulating polariton propagation~\cite{Kim:2017,Dai:2019,Lee:2020,Herzig:2024}. Gradual changes in the substrate can locally modify the polariton wavelength and momentum~\cite{Zheng:2022a,Yu:2023}, potentially enabling functionalities such as wavefront shaping and nanoscale focusing. 

Experimental demonstration of polariton nanofocusing has been primarily limited to the in-plane nanofocusing, driven by its potential applications in on-chip flat-optics, enhanced light-matter interaction for sensing, and its inherently simpler realisation. Most nanofocusing schemes rely on the intrinsic in-plane optical anisotropy of HPhPs in $\alpha$-phase molybdenum trioxide ($\alpha$-MoO$_3$) ~\cite{Zhang:2021, Duan:2021, Martin-Sanchez:2021, Zheng:2022b, Qu:2022}, though very recently, hBN too, has been engineered to perform similarly~\cite{Borodin:2026}. 

In contrast, a lateral nanofocusing design, where confinement occurs along the out-of-plane direction has received comparatively little attention despite its relevance for the same scope of applications. To the best of our knowledge, the lateral nanofocusing of HPhPs has been experimentally demonstrated only once -- in a tapered hBN flake where the progressive reduction of the crystal thickness leads to a gradual compression of the modes propagating towards the taper apex~\cite{Nikitin:2016}. Another concept has been proposed theoretically for acoustic graphene plasmons~\cite{Alonso-Gonzalez:2017}, predicting that a monotonically narrowing gap between graphene and a metallic substrate can progressively increase the plasmon momentum, resulting in strong field compression and lateral nanofocusing~\cite{Morozov:2019, Voronin:2020}. These theoretical works highlight the spatially varying polariton-substrate separation as a powerful mechanism to control polariton momentum. However, implementing a smoothly varying substrate geometry, let alone doing so in a reproducible and scalable manner, has been extremely challenging so far.

Here, we employ a platform that enables continuous and controllable spatial modulation of the hBN-substrate separation using a sinusoidally corrugated azopolymer film -- a highly versatile and scalable platform, well suited for a diverse range of photonic applications both in near- and far-field~\cite{Lim:2021}. For example, we recently demonstrated a new type of polaritonic crystal that is based on the same platform~\cite{Menabde:2025a, Menabde:2026}. In this work, we deposit gold (Au) on the corrugated azopolymer substrate and show that the resulting structure enables a precise and local engineering of HPhPs momentum in hBN via a quasi-adiabatic variation of their effective index provided by smoothly varying gap between the hBN and Au, demonstrating a $\sim2.7$ times tuning of the in-plane HPhP momentum $q$. Such local tailoring of polaritons opens new pathways toward mid-infrared optical rulers, polaritonic heat-transfer devices, tunable micro-electromechanical systems, and integrated polaritonic circuitry.

Furthermore, we leverage the gradual variation of the gap across the corrugation pattern to demonstrate, for the first time, continuous compression and decompression of the propagating polaritonic modes in a uniform vdW crystal driven solely by substrate geometry. We observe an approximately 2.5 times change of the HPhP wavelength -- in excellent agreement with both local control experiments and theoretical calculations -- demonstrating lateral nanofocusing of propagating polaritons. Lateral nanofocusing would be important in applications where adiabatic momentum adjustment is necessary, such as for example, highly efficient polaritonic couplers.

\section{Results}
\begin{figure}[htbp]
\centering
\includegraphics[]{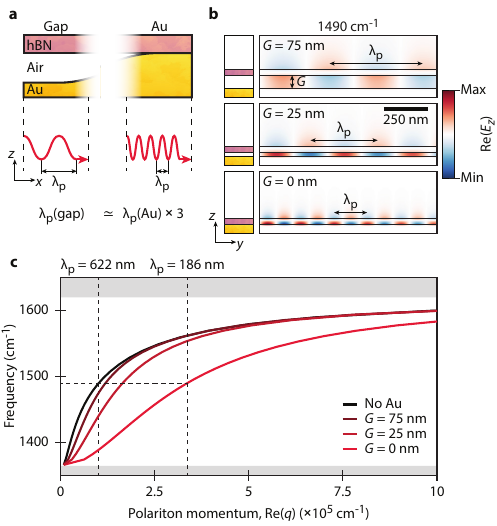}
\caption
{\textbf{Concept of substrate-tunable phonon polaritons.} \textbf{a}, Schematic of HPhP wavelength control through hBN-substrate separation. The structure on the left has hBN separated by a gap to the Au substrate resulting in a polariton wavelength of $\lambda_\mathrm{p}(\mathrm{gap})$. The structure on the right has hBN in direct contact with the Au substrate resulting in a polariton wavelength of $3\lambda_\mathrm{p}(\mathrm{Au}) \simeq \lambda_\mathrm{p}(\mathrm{gap})$.
\textbf{b}, Numerically calculated $\mathrm{Re}(E_z)$ field profiles at a frequency of 1490\,cm$^{-1}$ for 33\,nm thick hBN suspended over Au with gaps of 75, 25, and 0\,nm.
\textbf{c}, Theoretical dispersion relations of freely suspended hBN and hBN suspended over Au with gaps of 75, 25, and 0\,nm -- the hBN is 33\,nm thick. The white shaded area indicates the upper Reststrahlen band where HPhPs are supported.
}
\label{fig1}
\end{figure}

We begin with theoretically introducing a concept of controlling the wavelength of HPhP modes through engineering of the hBN-substrate separation. In hBN, HPhPs exist within the spectral Reststrahlen bands where the in-plane and out-of-plane permittivities have the opposite sign, resulting in hyperbolic dispersion and strongly confined polaritonic modes. In the following, we focus on the upper Reststrahlen band where HPhPs exhibit comparatively long propagation lengths and are well suited for near-field imaging experiments. Figure~\ref{fig1}a illustrates an idealised hBN/air/Au system. On the left, hBN is separated from Au by a significantly wide air gap, effectively acting as a suspended hBN. On the right, hBN is in direct contact with the Au substrate. In this case, polaritons in hBN couple to their mirror image in the metal, forming so-called image HPhPs with wavelengths more than three times shorter than in suspended hBN~\cite{Ambrosio:2018,Lee:2020,Menabde:2022b}. 

In order to get an insight into the sensitivity of polaritons to the distance bewteen the hBN and the metal, we examine the HPhP wavelength as a function of the gap width $G$ (Fig.~\ref{fig1}b,c). For the largest gap ($G=75$\,nm), the mode closely resembles the fundamental mode in a free-standing hBN~\cite{Menabde:2022a} with the anti-symmetric $E_z$ field profile and confined primarily outside the flake. The mode's dispersion at large gap (burgundy curve in Fig.~\ref{fig1}c) only slightly deviates from the free-standing case (black curve in Fig.~\ref{fig1}c). As the gap narrows ($G=25$\,nm), the HPhP field starts to experience screening by Au, increasingly penetrates the hBN, and concentrates within the gap, indicating the gradual evolution into an image mode~\cite{Menabde:2022a}. This is accompanied by a noticeable compression of the polariton's wavelength and a marked increase of its momentum (dark red curve in Fig.~\ref{fig1}c). The image mode fully manifests when the hBN is in direct contact with Au ($G=0$\,nm), resulting in the strong field localisation within the hBN and a substantially reduced HPhP wavelength, leading to a more than threefold increase of the polariton momentum around the middle of the Reststrahlen band (bright red curve in Fig.~\ref{fig1}c). This pronounced shift in dispersion underscores the exceptional tunability enabled by substrate engineering, stemming from the continuous evolution of the fundamental mode in suspended hBN into the strongly screened image HPhP in the area where hBN is on Au~\cite{Menabde:2022b}.

\begin{figure*}[htbp]
\centering
\includegraphics[]{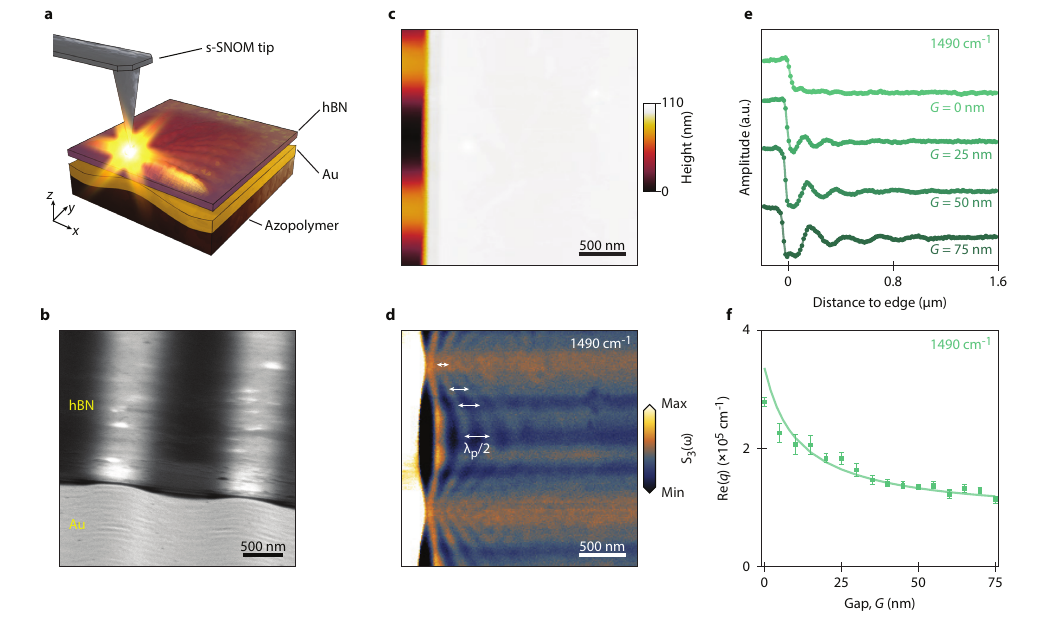}
\caption
{\textbf{Experimental structure and real-space nanoimaging of substrate-tunable phonon polaritons.} \textbf{a}, Experimental structure: azopolymer sinusoidally corrugated surface with conformally deposited Au and an hBN flake on top, probed by a metallic s-SNOM tip. 
\textbf{b}, Scanning electron microscopy image of the sample. 
\textbf{c}, Surface topography of the hBN edge obtained by the AFM.
\textbf{d}, Near-field image of the area shown in \textbf{c}, measured by s-SNOM at the excitation frequency of 1490\,cm$^{-1}$.
\textbf{e}, Line scans extracted from amplitude of near-field signal in (\textbf{d}) at gaps between hBN and Au $G=0$, 25, 50, and 75\,nm. The line scans are offset for ease of comparison, but not normalised.
\textbf{f}, Measured (markers) and theoretical dispersion (lines) of $\mathrm{Re}(q)$ as a function of gap between hBN and Au.
}
\label{fig2}
\end{figure*}

Our experimental approach is based on mapping the polaritonic near-field with a scattering-type scanning near-field optical microscope (s-SNOM)~\cite{Hillenbrand:2025} (Fig.~\ref{fig2}a). As has been shown in literature, the near-field interference fringes close to hBN edges emerge from the superposition of the tip-launched and the edge-reflected HPhP modes~\cite{Dai:2014, Dai:2017, Menabde:2022b}. The periodicity of the near-field fringes reveals the polariton wavelength, $\lambda_\mathrm{p}$, as it is equal to $\lambda_\mathrm{p}/2$, from which the in-plane wavevector $\mathrm{Re}(q) = 2\pi/\lambda_\mathrm{p}$ can be derived, enabling a direct extraction of the polariton dispersion. We probe polaritons in the middle of the hBN's upper Reststrahlen band at the excitation frequencies between 1450 and 1510\,cm$^{-1}$ where HPhPs exhibit strong confinement while maintaining sufficient propagation length, which is necessary for imaging of clear near-field interference pattern. The probed hBN edge in our samples is always oriented perpendicular to the substrate corrugation (Fig.~\ref{fig2}a)

Building upon the sensitivity of the HPhP momentum to the air gap, we employ a sinusoidally corrugated Au substrate for hBN (Fig.~\ref{fig2}a). Such a substrate is fabricated using a wafer-scale azopolymer film patterned by holographic inscription~\cite{Lim:2021}, followed by a thermal deposition of thin ($\sim30$\ nm) Au layer. Our corrugated mirrors have a period of $\simeq1500$\,nm and depth modulation of $\simeq75$\,nm (see Methods). Exfoliated hBN flakes of desired thickness are transferred onto the mirrors by a dry transfer technique. In the following discussion, we focus on a representative sample with hBN thickness of $d\simeq33$\,nm. Additional measurements of a thinner ($d\simeq15$\,nm) and thicker flake ($d\simeq88$\,nm) show qualitatively identical behaviour and are presented in Supplementary information section~S1. Oblique-view scanning electron microscopy (SEM; Fig.~\ref{fig2}b) and atomic force microscopy (AFM; Fig.~\ref{fig2}c) confirm that the hBN rests flat atop the corrugated substrate without conforming to its topography. Thus, the varying hBN-Au separation forms a position-dependent environment that locally modulates $q$ and shapes their wavefront across the structure.

Near-field imaging of the hBN edge at an excitation frequency of 1490\,cm$^{-1}$ clearly reveals a spatial modulation of the polaritonic wavelength above the sinusoidally corrugated Au surface (Fig.~\ref{fig2}d). As mentioned above, the near-field interference fringes directly visualise $\lambda_\mathrm{p}$, revealing its variation along the edge and confirming that the corrugated substrate locally modifies $q$. The simultaneously acquired sample topography (Fig.~\ref{fig2}c) shows that $\lambda_\mathrm{p}$ correlates with the substrate geometry. 

To quantify the gap-induced tuning of the HPhP dispersion, we analyse the interference pattern in Fig.~\ref{fig2}d by extracting the line scans of the near-field amplitude perpendicular to the hBN edge (Fig.~\ref{fig2}e), while the local gap height is determined from the correlated AFM maps. Measurements of the sample at different excitation frequencies, performed at 10\,cm$^{-1}$ increment, show consistent behaviour and are presented in Supplementary information section~S2. Then, we perform a fast Fourier transform (FFT) of each line scan and fit the resulting spectrum with that of a damped harmonic oscillator. The peak of the fitted spectrum yields the $\mathrm{Re}(q)$ with high accuracy. As the FFT-based analysis of polaritonic fringes is well established in the literature~\cite{Jang:2024}, we omit technical details here and provide a full procedure description in Supplementary information section~S3.

The experimentally obtained dependency of $\mathrm{Re}(q)$ on $G$ reveals the full extent of the gap-induced dispersion tuning (Fig.~\ref{fig2}f, data points), also showing an excellent agreement with theoretical prediction (Fig.~\ref{fig2}f, solid line). The momentum decreases exponentially with increasing the gap size, approaching the free-standing limit at $G=75$\,nm. Overall, we observe a $\sim2.7$ times tuning of $\mathrm{Re}(q)$ between the in-contact and suspended regimes, demonstrating that the corrugated metallic substrate enables a strong spatial control over the HPhPs propagation.

\begin{figure}[tbp]
\centering
\includegraphics[]{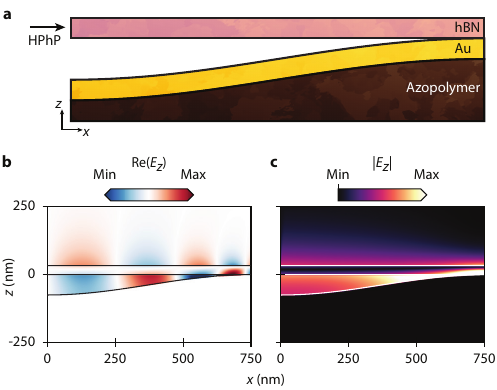}
\caption
{\textbf{Concept of lateral nanofocusing.} \textbf{a}, Schematic of a polaritonic taper formed by a corrugated substrate. HPhPs propagate from a region with large hBN-substrate separation towards a region of direct contact between hBN and Au.
\textbf{b,c}, Simulated real part (\textbf{b}) and magnitude (\textbf{c}) of the HPhP $\mathrm{E}_\mathrm{z}$ field, at a frequency of $1490\,\mathrm{cm}^{-1}$, showing progressive wavelength compression and field confinement along the taper.
}
\label{fig3}
\end{figure}

The demonstrated high tunability of $\mathrm{Re}(q)$ opens avenues for even more versatile spatial control: perpendicular to the corrugation, the substrate forms a naturally tapered polaritonic waveguide (Fig.~\ref{fig3}a). Considering the HPhPs propagating in this direction, we expect that their momentum would smoothly change as they propagate. If $G$ along the propagation direction decreases, the corresponding increase of polariton momentum will lead to a progressively larger field confinement, thus resulting in lateral nanofocusing of the propagating polaritons. We note that, to the best of our knowledge, lateral nanofocusing has been demonstrated only once: using a naturally formed taper at the edge of an exfoliated hBN flake~\cite{Nikitin:2016}. However, such taper geometries are difficult to reproduce, as exfoliated hBN crystals typically yield sharp, abrupt edges, and nanopatterning of hBN would degrade its surface quality compared to the pristine exfoliated crystals.

Our full-wave simulations reveal a pronounced compression of the polariton wavelength as the mode propagates along the taper (Fig.~\ref{fig3}b,c; see Methods). Despite the relatively steep and nonlinear substrate gradient, the mode evolution remains largely adiabatic, with minimal back-reflection or scattering into the higher-order modes. Near the apex, our simulations display a strong field enhancement characteristic of the image HPhP mode, with the $E_z$ field becoming highly localised within the hBN and the narrowing gap. This behaviour confirms that the spatially varying gap acts as an effective polaritonic taper capable of compressing and focusing propagating HPhPs, without the need to pattern a vdW crystal.

\begin{figure*}[htbp]
\centering
\includegraphics[]{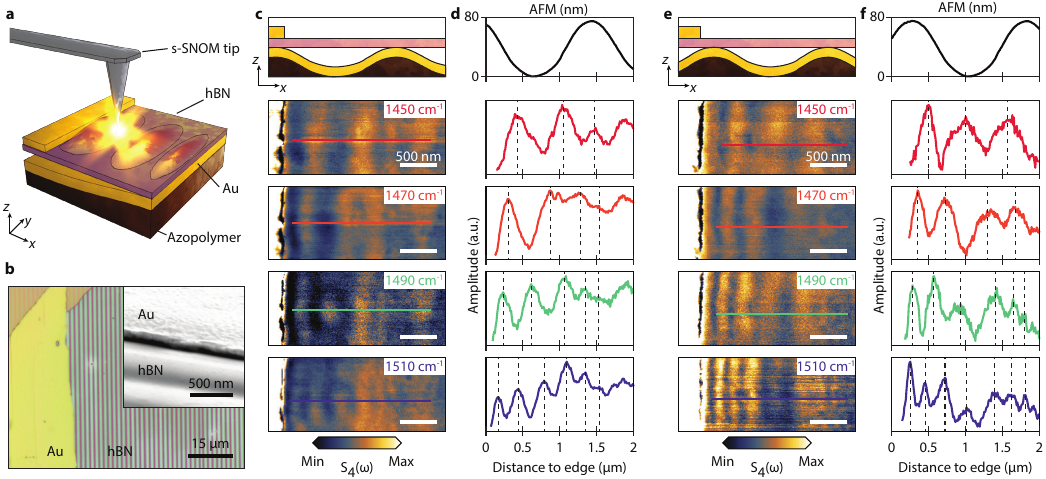}
\caption
{\textbf{Systematic measurement of lateral nanofocusing.} \textbf{a}, Experimental structure: azopolymer sinusoidally corrugated surface with conformally deposited Au and an hBN flake on top; HPhPs are launched by a gold piece carefully placed on top of hBN and probed by a metallic AFM tip.
\textbf{b}, Bright-field optical miscroscopy image of the sample; inset shows an oblique angle scanning electron microscopy image.
\textbf{c}, Near-field maps measured by s-SNOM at excitation frequencies between 1450\,cm$^{-1}$ and 1510\,cm$^{-1}$ with increment an of 20\,cm$^{-1}$. The sketch above illustrates the heterostructure, showing the position of the Au flake and the underlying topography of the corrugated surface. \textbf{d}, Line scans extracted from the s-SNOM measurements in (\textbf{c}). The dashed black lines indicates peak fringe peak positions. The AFM topography of the corrugated surface (north of the scan area) is shown in the top panel. \textbf{e,f}, Same as (\textbf{c}) and (\textbf{d}), but for a different position on the sample.
}
\label{fig4}
\end{figure*}

In order to experimentally probe the lateral nanofocusing, we must launch HPhPs perpendicular to the substrate corrugation, which requires a linear edge scatterer on top of hBN (Fig.~\ref{fig4}a). To that end, we develop a transfer technique based on thin-film delamination. An 80\,nm Au film is first deposited on a polymethyl methacrylate (PMMA)-coated silicon substrate. Dissolving the PMMA in acetone weakens adhesion, allowing the Au film to be collected with a PDMS stamp and mechanically "exfoliated" into strips that can be transferred onto the hBN (Fig.~\ref{fig4}b; see Methods and Supplementary information~S4 for details). This approach enables us to carefully control the position of the metallic edge in the sample, providing an efficient launching of HPhPs in the desired direction.

With the Au launcher in place, we can investigate the propagation and nanofocusing of HPhPs. First, we position the scattering edge approximately 100\,nm upstream of a corrugation apex to efficiently launch HPhPs in the suspended hBN (top panel in Fig.~\ref{fig4}c). Importantly, the metallic-edge-launched HPhPs form near-field fringes by interfering with a quasi-uniform excitation beam, thus the separation of fringes is equal to $\lambda_\mathrm{p}$ ~\cite{Jang:2024}. s-SNOM images reveal direct evidence of lateral nanofocusing: the polariton wavelength is gradually compressed as the mode propagates towards the opposite side of the valley (Fig.~\ref{fig4}c). Note that the background contributions from the corrugated Au substrate -- prominent in Fig.~\ref{fig2}d -- is largely suppressed by analysing the fourth-order demodulation harmonic ($n=4$) of the near-field signal~\cite{Ocelic:2006}. Despite the higher-order detection, the near-field fringes remain well-resolved, enabling clear visualisation of the nanofocusing behaviour.

To quantify the observed mode compression, we analyse the near-field line scans across the tapered region (coloured lines in Fig.~\ref{fig4}c), averaging them over 11 pixels-wide ribbons. At all frequencies from 1450 to 1510\,cm$^{-1}$, we observe a clear compression of the polariton wavelength as the mode propagates from the launcher (Fig.~\ref{fig4}d). The effect is most pronounced at $\omega=1490\,\mathrm{cm}^{-1}$: the peak-to-peak fringe spacing decreases from $\lambda_\mathrm{P} \simeq 451$\,nm to $\simeq 180$\,nm, corresponding to a 2.5-fold reduction (see Supplementary information~S5 for additional peak-to-peak fringe spacing analysis). The frequency at which maximum compression occurs depends on the hBN thickness, its dielectric response, the substrate topography, and the placement of the Au edge.

We further investigate how the focusing dynamics depends on the launching conditions by examining another region of the sample where the Au edge is positioned $\sim150$\,nm downstream of the corrugation peak (top panel in Fig.~\ref{fig4}e). In this case, HPhPs initially experience lateral compression, but subsequently undergo decompression (defocusing), followed by another focusing as they propagate across the sinusoidal surface (Fig.~\ref{fig4}e). This phenomenon is visible as a spatial modulation of fringe spacing in the extracted near-field line scans (Fig.~\ref{fig4}f) for $\omega=1470-1510\,\mathrm{cm}^{-1}$. The effect is especially pronounced at $\omega=1510\,\mathrm{cm}^{-1}$ where the mode first expands by a factor of $\sim1.9$ before refocusing by $\sim2$ times, clearly demonstrating a never-before-seen reversible control of polariton confinement through substrate geometry. Additional measurements of a thinner ($d\simeq15$\,nm) hBN flake show qualitatively identical behaviour and are presented in Supplementary information section~S1.

\begin{figure}[htbp]
\centering
\includegraphics[]{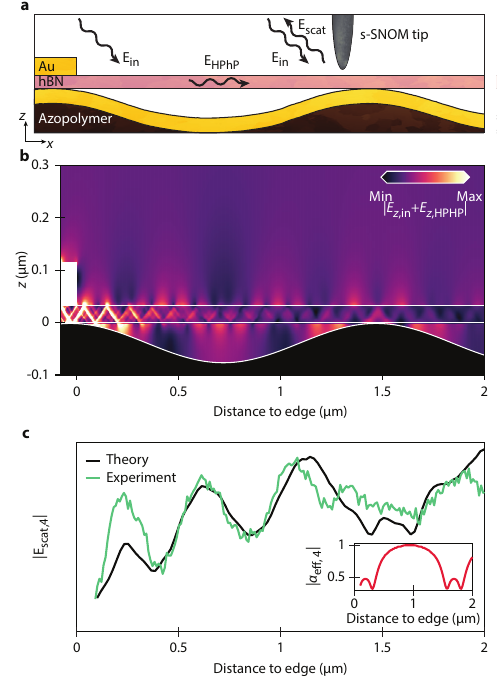}
\caption
{\textbf{Lateral nanofocusing of HPhPs.} \textbf{a}, Schematic of simulation configuration. Incident light on the Au edge launches HPhPs that propagate along the corrugated structure, while the near-field signal is captured at a finite distance above the hBN surface to emulate the s-SNOM tip.
\textbf{b}, Numerical calculation of the $|E_{z,\mathrm{in}}+E_{z,\mathrm{HPhP}}|$ field at the frequency of $1490\,\mathrm{cm}^{-1}$. The field is saturated to improve readability.
\textbf{c}, Field amplitude, at the frequency of $1490\,\mathrm{cm}^{-1}$, from experiment (green) and numerically calculated s-SNOM signal (black), given by $E_\mathrm{scat,4}\propto \alpha_\mathrm{eff,4}(E_{z,\mathrm{in}}+E_{z,\mathrm{HPhP}})$. The inset shows the absolute value of $\alpha_\mathrm{eff,4}$ as a function of position, highlighting a reduced tip-sample coupling near a taper apex.
}
\label{fig5}
\end{figure}

Although the experiments demonstrate the wavelength compression of propagating HPhPs, our measurements do not reveal an expected increase in near-field signal amplitude at the focusing areas. Such field enhancement would be particularly desirable for applications in optical circuitry, nonlinear optics, or sensing, where strong local fields are essential. To investigate the apparent absence of the near-field enhancement, we turn to full-wave simulations of the near-field probing (see Methods). Figure~\ref{fig5}a displays the simulation model that replicates the experiment in Figs.~\ref{fig4}c,d, where incident light scatters at the Au edge and launches the HPhP modes. 

Figure~\ref{fig5}b displays the calculated magnitude of the total electric field's $z$-component, $|E_{z,\mathrm{in}}+E_{z,\mathrm{HPhP}}|$, at a frequency of $\omega=1490\,\mathrm{cm}^{-1}$. Although the shart Au edge initially launches infinitely many higher-order modes in hBN -- producing the characteristic ray-like field pattern -- these modes decay quickly. Only the fundamental HPhP mode propagates across the full corrugation cycle, exhibiting a strong field concentration at the apex: this hot spot confirms efficient lateral nanofocusing.

To enable the quantitative comparison between our s-SNOM experiment and the simulation, we extract the scattered near-field from the simulation, $E_\mathrm{scat,4}$, at a distance of 65\,nm above the hBN, corresponding to the experimental conditions~\cite{Ocelic:2006}. Furthermore, to account for the tip-sample interaction, we calculate the effective polarisability of the tip, $\alpha_\mathrm{eff}$, using the well-established point-dipole model~\cite{Huber:2005}. Then, the simulated s-SNOM signal is given by $E_\mathrm{scat,4}\propto \alpha_\mathrm{eff,4}(E_{z,\mathrm{in}}+E_{z,\mathrm{HPhP}})$, where $\alpha_\mathrm{eff,4}$ corresponds to the polarisability at the fourth demodulation harmonic ($n=4$). Overall, the simulated signal accurately reproduces both the polariton wavelength and the field amplitude observed in s-SNOM measurements (Fig.~\ref{fig5}c). Importantly, the inset in Fig.~\ref{fig5}c displays the spatial dependence of $|\alpha_\mathrm{eff,4}|$, revealing that the near-field coupling is more than halved over the corrugation apex, which explains the absence of observable field enhancement in the experiment. Summarising, our results indicate that the substrate-engineered polaritonic tapers enable efficient lateral nanofocusing, while highlighting the importance of the tip-sample interaction for the correct interpretation of near-field measurements.

\section{Discussion}
In conclusion, we demonstrate a substrate-induced control of HPhP dispersion by leveraging a gilded sinusoidally corrugated surface to locally modulate the hBN-Au air gap. This approach enables precise in-plane wavefront engineering and reveals the rapid transition from HPhP modes in suspended hBN to highly confined image HPhPs in hBN on Au, with a nearly threefold wavelength compression. Beyond the local control, our platform supports lateral nanofocusing and defocusing of propagating modes, achieving a wavelength variation of around 2.5 times. These results establish substrate engineering as a powerful tool for tailoring polariton behaviour and open new possibilities for nanophotonic devices.

As previously suggested for the HPhPs in $\alpha$-MoO$_3$~\cite{Yu:2023}, the significant modulation of the polariton momentum suggests that hBN may serve as an optical ruler capable of resolving thickness or dielectric properties of ultra-thin and low-index materials that are otherwise inaccessible by conventional ellipsometry~\cite{Heiden:2025}. In addition, HPhPs in hBN on Au recently have been reported to be excellent at heat transfer~\cite{Hutchins:2025}, suggesting that a corrugated metallic substrate may prove useful in managing the heat. Furthermore, the strong dependence of the polariton wavelength on the gap size opens opportunities for tunable micro-electromechanical systems (MEMS)-based polaritonic metasurfaces in the mid-infrared, analogous to recent advances in tuning plasmonic metasurfaces operating at near-infrared frequencies~\cite{Meng:2021}.

Furthermore, this work represents only the second reported instance of lateral nanofocusing of propagating polaritons~\cite{Nikitin:2016}, and the first to achieve it through substrate modulation rather than tapering the waveguiding material itself. By shaping the underlying surface, we demonstrate a practical and reproducible strategy for nanofocusing, producing highly confined polaritons launched from low-momentum modes. This method offers a practical route for wavelength management in polaritonic systems and may facilitate efficient coupling between different polariton types~\cite{Nikitin:2014} -- a key requirements for the development of future integrated polaritonic circuitry.

\section{Methods}
\subsection{Sample preparation}
We prepared a 3 weight percent solution of poly(disperse red 1 methacrylate) (pDR1m; $Mw$=3 kDa and polydispersity index (PDI)=1.1--1.2) by dissolving this polymer in 1,1,2-trichloroehtane. The pDR1m was synthesised in-house, and, as we previously reported,~\cite{Lim:2021} its $Mw$ and PDI were optimised to enable the most efficient formation of the Fourier surface. We spin-coat this solution onto a $2\times2\,\mathrm{cm}^{2}$ Si wafer at 1000\,rpm for 40\,s, resulting in an azopolymeric thin film with the thickness of 150\,nm. Then, using a continuous wave diode laser (Light House Sprout) with a wavelength of 532\,nm, we mixed two beams with left- and right-hand circular polarizations and thereby generated polarisations interference pattern (PIP). The mere illumination of this PIP onto an azopolymeric thin film allowed us for inscribing a single-harmonic Fourier surface. The intensity of each beam was 150\,mW/cm$^2$. Fabrication of the Fourier surface with a period of 1500\,nm and a modulation height of 75\,nm required an inscription time of 8\,min and a beam incident angle of 20\,degrees. Using thermal evaporation deposition at a rate of 0.2\,Å\,s$^{-1}$ with a vacuum pressure of $\approx5\times10^{-7}$\,Torr, we deposited an approximately 30\,nm thick Au layer on the azopolymeric Fourier surface.

To transfer hBN to the gilded Fourier surface, we mechanically exfoliate epitaxial solidification-grown hBN, commercially sourced from 2D semiconductors USA, using low adhesion Nitto SPV224 tape (blue tape). We exfoliate the original hBN crystal until the tape is visibly covered, then switch to a fresh piece of tape for a last round of exfoliation on a polydimethylsiloxane (PDMS) stamp, commercially available from GELPAK.

We select the hBN flakes on the PDMS stamp by comparing their visual contrast, and the chosen flake is transferred using a manual transfer stage at a temperature of $80^\circ\mathrm{C}$.

To create a scattering edge for probing lateral nanofocusing, we deposit 80\,nm Au on a PMMA-coated silicon substrate using thermal evaporation deposition at a rate of 0.5\,Å\,s$^{-1}$ with a vacuum pressure of $\approx5\times10^{-7}$\,Torr. After dissolving the PMMA in acetone we retrieve the Au film with a PDMS stamp, mechanically "exfoliate" it into fragments, and transfer the selected Au strip using a manual transfer stage at a temperature of $80^\circ\mathrm{C}$ (see Supplementary information~S4 for further details).

\subsection{s-SNOM measurements}
We perform near-field imaging using a commercial s-SNOM (neaSNOM from attocube systems AG, formerly Neaspec) coupled to a tunable quantum cascade laser (MIRcat from Daylight Solutions). To probe the propagating phonon-polaritons, we employ Pt/Ir-coated AFM tips (ARROW-NCPt-50 from NanoWorld) operating in tapping mode, with a tapping frequency of approximately 270\,kHz and an oscillation amplitude between 60 and 70\,nm. 

For image generation, we utilise the pseudo-heterodyne interferometric detection module, where background-free interferometric signals are obtained by demodulating the scattering amplitude (the output signal of the s-SNOM) at different $n$-order harmonics of the tapping frequency. For the systematic dispersion dependence on gap measurements, we select the $n=3$ harmonic to ensure near-field images with low background noise -- some of the background signal is, however, still present and we thus employ the $n=4$ harmonic for further background reduction in the polariton focusing experiment, albeit with a lower signal-to-noise ratio.

Additionally, during the s-SNOM raster scans, of the surface, we simultaneously acquire topography data for the scanned area, thereby providing precise thickness measurements of the hBN flakes and the gap between hBN and Au, ensuring consistency across measurements.

\subsection{Theoretical modelling and simulations}
Full-wave numerical simulations corresponding to the profiles shown in Figure~\ref{fig1}b, Figure~\ref{fig3}b,c, and Figure~\ref{fig5}b were performed using the finite-element method in the frequency domain in 2D space, implemented in COMSOL Multiphysics. The  gilded Fourier surface was modelled as a perfect electric conductor (PEC). The simulated s-SNOM signal $E_\mathrm{scat,4}$ presented in Figure~\ref{fig5}c is given by
$
E_\mathrm{scat,4} \propto \alpha_\mathrm{eff,4}(E_{z,\mathrm{in}} + E_{z,\mathrm{HPhP}}),
$
where the effective polarisability $\alpha_\mathrm{eff,4}$ was evaluated using an approximate analytical expression derived in Ref.~\cite{Kim:2015}. This approach captures the fourth-harmonic demodulated signal in the point-dipole model for s-SNOM, allowing direct comparison with experiment.

\section{Data availability}

The data that underlie the findings of this study are available from the corresponding authors upon reasonable request.

%\section{References}
\bibliographystyle{apsrev4-2}
\bibliography{references}

\section{Acknowledgments}

We thank M.-K.~Seo for lending us equipment.
This work was supported by the National Research Foundation of Korea (NRF) grants funded by the Korea government (MSIT) (2022R1A2C2092095, RS-2024-00414119, RS-2024-00416583, RS-2024-00340639). 
S.~L. acknowledges funding from the National Research Foundation of Korea (grant number: RS-2022-NR068141). This research was also supported by a grant of the Korea-US Collaborative Research Fund (KUCRF), funded by the Ministry of Science and ICT and Ministry of Health \& Welfare, Republic of Korea (grant number: RS-2024-00468463) and by Korea University grant.
A.~Y.~N. acknowledges the Department of Science, Universities and Innovation of the Basque Government (grant PIBA-2023-1-0007 and the IKUR Strategy) and the Spanish Ministry of Science and Innovation (grant PID2023-147676NB-I00).
P.~A.-G. acknowledges support from the European Research Council under Consolidator grant no. 101044461, TWISTOPTICS, the Spanish Ministry of Science and Innovation (State Plan for Scientific and Technical Research and Innovation grant number PID2019-111156GB-I00) and Agencia SEKUENS (Asturias) under grant UONANO IDE/2024/000678 with the support of FEDER funds.

\section{Author contributions statement}

J.~T.~H. and M.~S.~J. conceived the overall idea.  
Y.~L. and H.~N. fabricated the Fourier surface with supervising by S.~L.
J.~T.~H. assembled the samples with hBN and gold scatterers.
J.~T.~H. performed the scanning near-field experiments.
J.~T.~H., H.~T., and A.~Y.~N performed the theoretical analysis with contributions from S.~G.~M. and P.~A.-G.
All authors participated in the analysis of the data and the writing of the manuscript.

\section{Competing interests statement}

The authors declare no competing interests.

\end{document}

%-----TITLE-----
\title{SUPPLEMENTRAY INFORMATION\\
Local control and lateral nanofocusing of hyperbolic phonon polaritons}

%-----AUTHORS AND AFFILIATIONS-----
\author{Jacob~T.~Heiden\,\orcidlink{0000-0003-4505-6107}}
\affiliation{School of Electrical Engineering, Korea Advanced Institute of Science and Technology (KAIST), Daejeon 34141, Korea}

\author{{Haozhe~Tong},\orcidlink{0009-0008-9053-9312}}
\affiliation{Donostia International Physics Center (DIPC), Donostia-San Sebastián 20018, Spain}
\affiliation{Universidad del Pais Vasco/Euskal Herriko Unibertsitatea (UPV/EHU), Donostia-San Sebastián 20018, Spain}

\author{Yongjun~Lim}
\affiliation{Department of Biomicrosystem Technology, Korea University, Seoul 02841, Korea}

\author{Heerin~Noh}
\affiliation{KU-KIST Graduate School of Converging Science and Technology, Korea University, Seoul 02841, Korea}

\author{Pablo~Alonso-Gonz\'{a}lez\,\orcidlink{0000-0002-4597-9326}}
\address{Department of Physics, University of Oviedo, Oviedo 33006, Spain}
\address{Center of Research on Nanomaterials and Nanotechnology, CINN (CSIC-Universidad de Oviedo), El Entrego 33940, Spain}

\author{Alexey.~Y.~Nikitin\,\orcidlink{0000-0002-2327-0164}}
\address{Donostia International Physics Center (DIPC), Donostia-San Sebastián 20018, Spain}
\address{IKERBASQUE, Basque Foundation for Science, Bilbao 48013, Spain}

\author{Seungwoo~Lee\,\orcidlink{0000-0002-6659-3457}}
\affiliation{Department of Biomicrosystem Technology, Korea University, Seoul 02841, Korea}
\affiliation{KU-KIST Graduate School of Converging Science and Technology, Korea University, Seoul 02841, Korea}
\affiliation{Department of Integrative Energy Engineering (College of Engineering), Korea University, Seoul 02841, Korea}

\author{Sergey~G.~Menabde\,\orcidlink{0000-0001-9188-8719}}
\thanks{\href{mailto:menabde@kaist.ac.kr}{menabde@kaist.ac.kr}; \href{mailto:jang.minseok@kaist.ac.kr}{jang.minseok@kaist.ac.kr}}
\affiliation{School of Electrical Engineering, Korea Advanced Institute of Science and Technology (KAIST), Daejeon 34141, Korea}
\email{menabde@kaist.ac.kr}\email{jang.minseok@kaist.ac.kr}

\author{Min~Seok~Jang\,\orcidlink{0000-0002-5683-1925}}
\thanks{\href{mailto:menabde@kaist.ac.kr}{menabde@kaist.ac.kr}; \href{mailto:jang.minseok@kaist.ac.kr}{jang.minseok@kaist.ac.kr}}
\affiliation{School of Electrical Engineering, Korea Advanced Institute of Science and Technology (KAIST), Daejeon 34141, Korea}
\email{menabde@kaist.ac.kr}\email{jang.minseok@kaist.ac.kr}

\maketitle

% %----- TABLE OF CONTENTS -----
\noindent{\textbf{\textsf{CONTENTS}}}\\ 
\twocolumngrid
\begingroup 
    \let\bfseries\relax % ToC sections are set in bold in RMP - we don't want that, so we relax \bfseries to nothing
    \deactivateaddvspace % removes excessive vertical spacing between ToC section lines
    \deactivatetocsubsections % removes subsections from ToC
    \tableofcontents
\endgroup
\onecolumngrid

\section{Results with different hBN thickness}
\label{sec:thickness}
As support to the measurements of dispersion dependence on gap in the main paper we also perform similar measurements on samples with hBN thickness $d\simeq 15$ and 88\,nm and depth modulation of $\simeq 40$ and 70\,nm, respectively (Fig.~\ref{si_fig_15nm}a and \ref{si_fig_88nm}a). The dispersion is extracted as described in Supplementary information section~\ref{sec:analysis}.

\begin{figure}[htbp]
\centering
\includegraphics[]{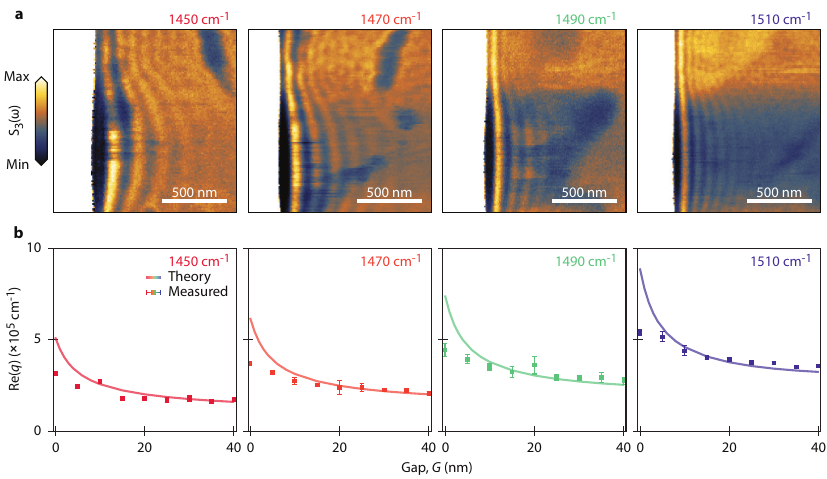}
\caption{
{\textbf{Systematic measurement of dispersion dependence on gap.} \textbf{a}, Spatial plot of near-field amplitude of a sample at frequency increments of 20\,cm$^{-1}$ between 1450\,cm$^{-1}$ and 1510\,cm$^{-1}$.
\textbf{b}, Measured (markers) and theoretical dispersion (lines) of $\mathrm{Re}(q)$ as a function of gap between hBN and Au.}}
\label{si_fig_15nm}
\end{figure}

\begin{figure}[htbp]
\centering
\includegraphics[]{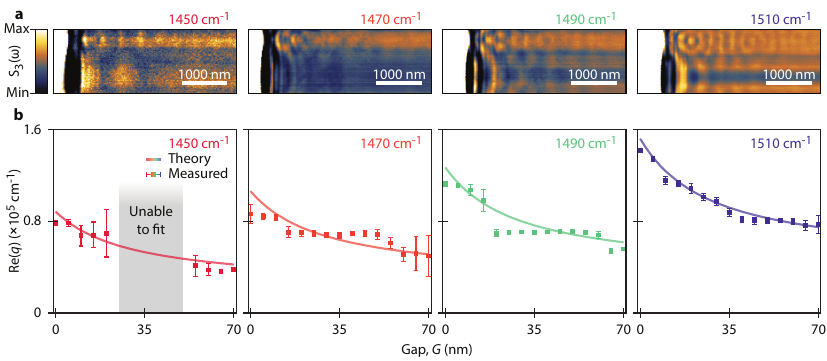}
\caption{
{\textbf{Systematic measurement of dispersion dependence on gap.} \textbf{a}, Spatial plot of near-field amplitude of a sample at frequency increments of 20\,cm$^{-1}$ between 1450\,cm$^{-1}$ and 1510\,cm$^{-1}$.
\textbf{b}, Measured (markers) and theoretical dispersion (lines) of $\mathrm{Re}(q)$ as a function of gap between hBN and Au.}}
\label{si_fig_88nm}
\end{figure}

Figure~\ref{si_fig_15nm}b and \ref{si_fig_88nm}b presents the $\mathrm{Re}(q)$ we extract from the measurements as a function of $G$, in fine overall agreement with the theoretical prediction across the full frequency range. The polariton momentum decreases exponentially with increasing gap size. However, in contrast to the $d \simeq 33$\,nm sample in the main text, the $d\simeq 15$\,nm sample exhibits noticeable deviations from theory when the hBN is in direct contact with the Au substrate. While it is not clear why this happens, one distinguishing feature of the thin hBN measurements is that the surface roughness of the underlying Au is visible in the near-field signal (Fig.~\ref{si_fig_15nm}a), suggesting that additional scattering may perturb the HPhP fringes. Additionally, the reduced thickness makes the HPhPs more susceptible to the influence of surface contamination, either on top of the hBN (as can be seen in figure~\ref{si_fig_15nm}a) or at the hBN-Au interface~\cite{Heiden:2025}, which may further affect the HPhP dispersion.

For the $d\simeq 88$\,nm sample, the tuning range of $\mathrm{Re}(q)$ is limited to a factor of about two (Fig.~\ref{si_fig_88nm}b), because the larger thickness prevents the phonon polaritons from approaching the free-standing hBN limit. Furthermore, extracting reliable values of $\mathrm{Re}(q)$ becomes increasingly difficult in the transition regime for thicker hBN. Over a broad range of gaps, the fringe spacing changes only weakly, and at 1450\,cm$^{-1}$ the data cannot be fitted for gap sizes between 25 and 50\,nm.

\newpage

\begin{figure}[H]
\centering
\includegraphics[]{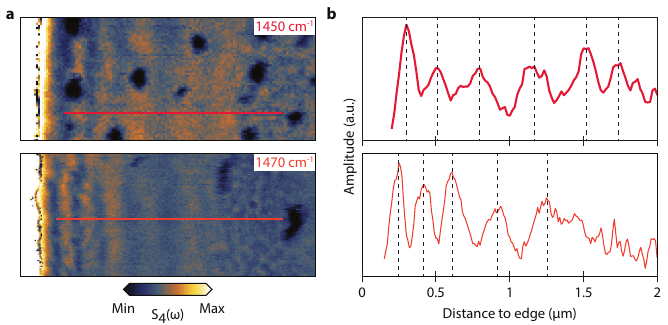}
\caption{
\textbf{Lateral nanofocusing.} \textbf{a}, Near-field maps measured by s-SNOM at excitation frequencies of 1450\,cm$^{-1}$ and 1470\,cm$^{-1}$. \textbf{b}, Line scans extracted from the s-SNOM measurements in (\textbf{a}). The dashed black lines indicates peak fringe peak positions.
}
\label{si_fig_15nm_focus}
\end{figure}

Lastly, we perform an extra focusing experiment,on the $d\simeq 15$\,nm sample, by exfoliating and transferring a Au strip on top of the hBN. By positioning the Au edge downstream of the corrugation peak, the launched HPhPs first propagate into a region of increasing hBN-Au separation, resulting in lateral defocusing, before being refocused as they experience the subsequent decreasing-gap region of the sinusoidal substrate (Fig.~\ref{si_fig_15nm_focus}). This behaviour is consistent with the defocusing described in the main text (Fig.~4e,f). Due to reduced signal strength (from the highly confined HPhPs in thin hBN) and the presence of surface contamination, this effect could be reliably observed only at excitation frequencies of 1450\,cm$^{-1}$ and 1470\,cm$^{-1}$.

\newpage

\section{Extra data for main sample}
\label{sec:extra_data}
To support the dispersion analysis presented in figure~2 of the main paper, we perform systematic near-field measurements on the same sample over a range of excitation frequencies from 1450\,cm$^{-1}$ and 1510\,cm$^{-1}$, in increments of 10\,cm$^{-1}$ (excluding at 1490\,cm$^{-1}$, which is shown in the main paper). The resulting spatial maps of the near-field amplitude are presented in figure~\ref{si_fig_extra}, from which the dispersion is extracted as described in Supplementary information section~\ref{sec:analysis}. The extracted dispersion shows good agreement with theoretical predictions across the full frequency range (Fig.~\ref{si_fig_extra}b), reconfirming that the gap-dependent tuning of HPhPs is robust and reproducible.

\begin{figure}[H]
\centering
\includegraphics[]{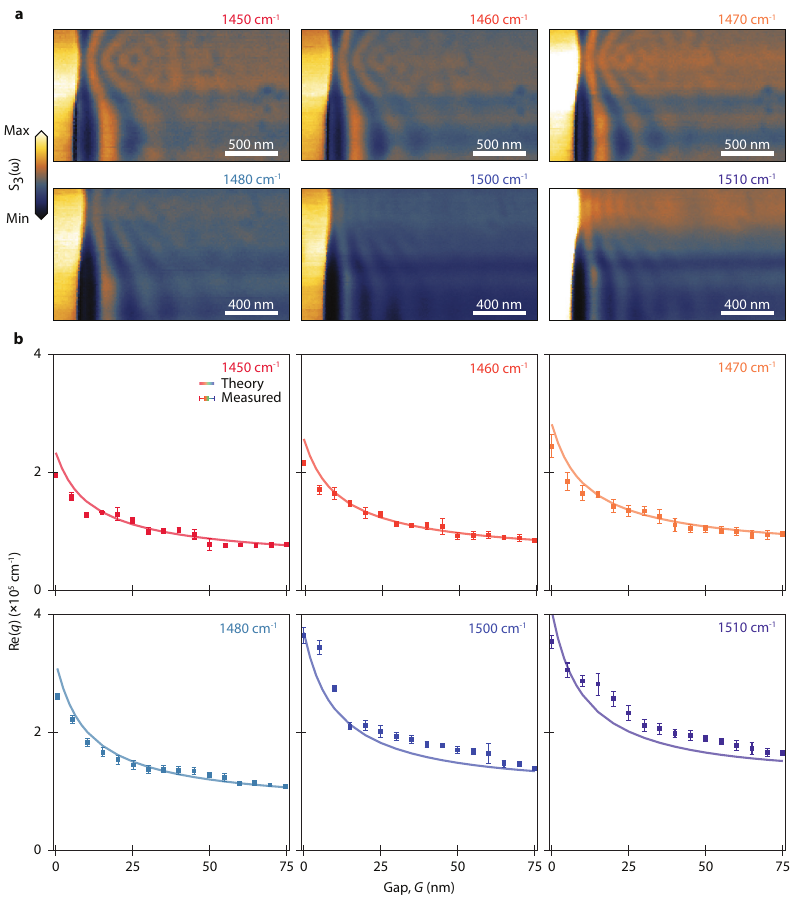}
\caption{
{\textbf{Systematic measurement of dispersion dependence on gap.} \textbf{a}, Spatial plot of near-field amplitude of a sample at frequency increments of 10\,cm$^{-1}$ between 1450\,cm$^{-1}$ and 1510\,cm$^{-1}$ (excluding at 1490\,cm$^{-1}$).
\textbf{b}, Measured (markers) and theoretical dispersion (lines) of $\mathrm{Re}(q)$ as a function of gap between hBN and Au.}}
\label{si_fig_extra}
\end{figure}

\newpage

\section{Near-field data analysis} 
\label{sec:analysis}
\begin{figure}[H]
\centering
\includegraphics[]{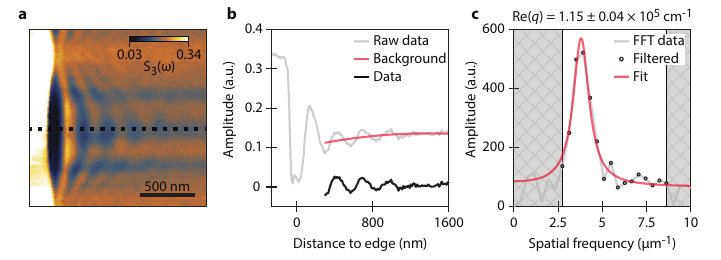}
\caption{
\textbf{Procedure for analysing near-field fringes.}
\textbf{a,} Spatial plot of the s-SNOM measurement of a sample at a frequency of 1490\,cm$^{-1}$. The black dashed line designate the line scan area at $G=75$\,nm. \textbf{b,} Original profile (grey), background (red), and result of removing the background (black). The edge position is assumed to be the first intensity dip; therefore, we truncate the signal at the second dip. \textbf{c,} Fourier spectrum (grey line), to which a rectangular frequency filter is applied (black circles) and the fit (red line).}
\label{SI_analysis}
\end{figure}

Here we describe our approach for characterising near-field data, using as an example the sample analysed in the main paper at an excitation frequency of 1490\,nm$^{-1}$ and a position where $G=75$\,nm (Fig.~\ref{SI_analysis}a). The methodology we employ is similar to the one presented by Jang \emph{et al.}~\cite{Jang:2024}, but is generalised for improved handling of multiple line scans and samples. 
\begin{enumerate}
\item \textit{Extract fringes.}
We perform line scans perpendicular to the hBN edge to extract data from the third harmonic demodulated s-SNOM plots (Fig.~\ref{SI_analysis}a,b). We truncate the signal at the first fringe, so that only the second fringe and beyond are included in the analysis. This exclusion helps eliminate parasitic near-field signals caused by strong scattering at the hBN edge.

\item \textit{Background removal.} 
The near-field data contains background signals resulting from the self-interference of incident light and the excitation of surface-plasmon-polaritons in the hBN/Au heterostructure~\cite{Jang:2024}. 
Here the background is modelled by polynomials with order $\le2$ and subtracted from the fringe data, setting the median to zero (Fig.~\ref{SI_analysis}b).

\item \textit{Fast Fourier transform and fit.} 
We apply a fast Fourier transform (FFT) to each line scan, transforming the data into the spatial-frequency domain (Fig.~\ref{SI_analysis}c). The tip-launched mode appears as a clear peak in the spectrum, providing the polariton wavevector $\mathrm{Re}(q)$. To ensure an accurate read, however, we first apply a rectangular frequency filter with a width of $1.5 \times \mathrm{FWHM}$ on the left side and $4 \times \mathrm{FWHM}$ on the right side. This filtering effectively removes the edge-launched component while retaining the tail of the tip-launched mode for fitting. 

Next, we fit the filtered spectrum. Since the $\left| \textbf{E} \right|$ profile of the launched polaritons follows a damped harmonic oscillator, we fit the Fourier-transformed data with:
\begin{equation}\label{eq:FFT_DHO}
F\left( s \right) = A \left( \frac{\gamma+\mathrm{i}s}
{\left( \gamma+\mathrm{i}s \right)^2 + k^2} \right) + C,
\end{equation}
where $\gamma$ is the polariton damping rate, $k$ is the spatial frequency, and $C$ is a constant. Fitting Eq.~\eqref{eq:FFT_DHO} to the Fourier spectrum yields $\mathrm{Re}(q) = 2\pi k$.
\end{enumerate}

\newpage

\section{Gold transfer} 
\label{sec:fabrication}
\begin{figure}[H]
\centering
\includegraphics[]{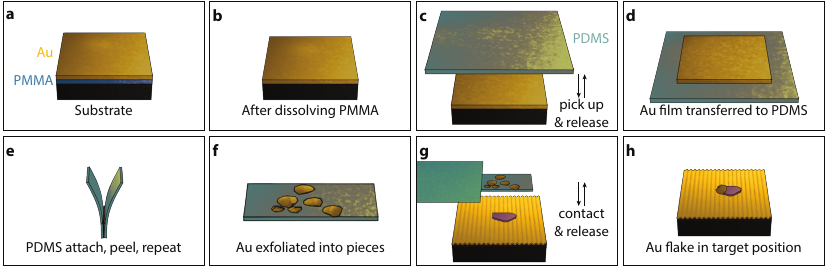}
\caption{
\textbf{Schematic of the Au exfoliation and transfer procedure.}
\textbf{a}, A sacrificial substrate with PMMA and Au on top. \textbf{b}, After dissolving PMMA in acetone, the Au film is weakly attached to the substrate. \textbf{c}, Retrieving the Au film from the substrate using a PDMS stamp. \textbf{d}, Au film is attached to the PDMS stamp. \textbf{e}, Exfoliation is done using two PDMS stamps gently pressed against the Au film and peeling multiple times. \textbf{f}, After several exfoliation cycles, a random distribution of Au pieces appears across the tape surface. \textbf{g}, The PDMS stamp is attached to a glass slide and the target Au piece is aligned over the target region on the substrate. Hereafter, the PDMS stamp is brought into contact with the substrate and slowly released again. \textbf{h}, The target Au piece has been transferred to the desired position on the substrate.
}
\label{si_fig_fabrication}
\end{figure}

In order to create a scattering edge capable of launching HPhPs perpendicular to the Fourier surface, we develop a Au transfer technique, outlined schematically in figure~\ref{si_fig_fabrication}. 

\begin{enumerate}
    \item We prepare the starting substrate by spin coating a polymethyl methacrylate (PMMA; Kayuka 950PMMA A4) sacrificial layer onto a Si wafer at 4krpm and bake at $180\,^\circ$C for 2\,min. The accuracy control is not expected to be critical. We then deposit the 80\,nm Au film by thermal evaporation at a rate of 0.5\,Å$s^{-1}$ with a vacuum pressure of $\approx5\times10^{-7}$\,Torr, leaving a Au/PMMA/Si substrate ready for further use.

    \item We dissolve the PMMA by immersing the sample in acetone for several hours and subsequently carefully rinse it in isopropanol and deionized water. This process releases the Au thin film that becomes very weakly attached to the sacrificial substrate (Fig.~\ref{si_fig_fabrication}b). The resulting film closely resembles leaf gold.

    \item  We retrieve the Au film from the substrate using a polydimethylsiloxane (PDMS) stamp, onto which the Au adheres upon contact (Fig.~\ref{si_fig_fabrication}c,d).

    \item We mechanically fragment the Au film into smaller pieces using a second PDMS stamp through repeated contact and separation, in a manner analogous to the exfoliation of two-dimensional van der Waals crystals (Fig.~\ref{si_fig_fabrication}e,f). To lower surface coverage and increase the separation between Au fragments, it may be rewarding to use a fresh PDMS stamp for each exfoliation step.

    \item To transfer a Au piece to a target location, we mount the receiving substrate on a vacuum stage that provides control over lateral position, rotation, in addition to temperature regulation. Due to the optical transparency of PDMS, both the Au piece and the substrate surface are simultaneously visible under a microscope, enabling accurate alignment. After alignment, we raise the substrate to bring it into contact with the PDMS stamp and gently press to ensure adhesion (Fig.~\ref{si_fig_fabrication}g). We then heat the substrate to $80\,^\circ$C to soften the PDMS and reduce its viscoelastic adhesion. Finally, we slowly retract the stage, causing the PDMS to peel away and leaving the Au fragment cleanly transferred onto the substrate surface (Fig.~\ref{si_fig_fabrication}h).
\end{enumerate}

\section{Peak-to-peak fringe spacing}
\label{sec:tables}
For completeness, we summarise the peak-to-peak fringe spacing associated with lateral focusing and defocusing for the investigated samples. We acquire these wavelengths by extracting the peak positions and then comparing the distance between peaks. Table~\ref{Table:focus1} lists the extracted polariton wavelengths for the $d \simeq 33$\,nm sample in the main text, in the focusing configuration (Fig.~4c,d), while table~\ref{Table:defocus1} lists the wavelengths in the defocusing configuration (Fig.~4e,f). Table~\ref{Table:defocus2} lists the wavelengths for the $d \simeq 15$\,nm sample (Fig.~\ref{si_fig_15nm_focus}). These values provide a rough reference for the observed spatial evolution of focused and defocused HPhPs.

\begin{table}[H]
\caption{
\textbf{Peak-to-peak HPhP focusing, extracted from figure~4d in the frequency range of 1450-1510\,cm$^{-1}$}
\label{Table:focus1}}
\centering
\begin{tabular}{lcccc}
\hline \hline
		& $1450\,\mathrm{cm}^{-1}$     	
        & $1470\,\mathrm{cm}^{-1}$    
        & $1490\,\mathrm{cm}^{-1}$    
		& $1510\,\mathrm{cm}^{-1}$		\\    
        & (nm)      & (nm)	        & (nm)          & (nm) 	 \\ \hline
\#1     & 621		& 571 			& 381 		    & 266    \\
\#2   	& 420		& 411  			& 451    		& 346    \\ 
\#3   	&     		& 251  			& 281    		& 306    \\ 
\#4   	& 		    &   			& 180    		& 253    \\ 
\#5   	& 		    &   			&          		& 186    \\ \hline \hline
\end{tabular}
\end{table}

\begin{table}[H]
\caption{
\textbf{Peak-to-peak HPhP defocusing, extracted from figure~4f in the frequency range of 1450-1510\,cm$^{-1}$}
\label{Table:defocus1}}
\centering
\begin{tabular}{lcccc}
\hline \hline
		& $1450\,\mathrm{cm}^{-1}$     	
        & $1470\,\mathrm{cm}^{-1}$    
        & $1490\,\mathrm{cm}^{-1}$    
		& $1510\,\mathrm{cm}^{-1}$		\\    
        & (nm)      & (nm)	        & (nm)          & (nm) 	 \\ \hline
\#1     & 501		& 371 			& 290 		    & 200    \\
\#2   	& 571		& 571  			& 361    		& 270    \\ 
\#3   	&     		& 381  			& 471    		& 290    \\ 
\#4   	& 		    &   			& 240    		& 381    \\ 
\#5   	& 		    &   			& 160    		& 230    \\ 
\#6   	& 		    &   			&        		& 190    \\ \hline \hline
\end{tabular}
\end{table}

\begin{table}[H]
\caption{
\textbf{Peak-to-peak HPhP defocusing, extracted from figure~\ref{si_fig_15nm_focus}b in the frequency range of 1450-1470\,cm$^{-1}$}
\label{Table:defocus2}}
\centering
\begin{tabular}{lcccc}
\hline \hline
		& $1450\,\mathrm{cm}^{-1}$     	
        & $1470\,\mathrm{cm}^{-1}$      \\    
        & (nm)      & (nm)	            \\ \hline
\#1     & 213		& 172 			    \\
\#2   	& 263		& 194  			    \\ 
\#3   	& 377    	& 310  			    \\ 
\#4   	& 367		& 326  			    \\ 
\#5   	& 234		&   			    \\ \hline \hline
\end{tabular}
\end{table}

\newpage

\section{Dielectric function of hBN} 
\label{sec:eps_hBN}
The dielectric function of our hBN crystals was carefully characterised in another paper~\cite{Heiden:2025}, using a combination of Raman spectroscopy and fitting of phonon polariton fringes extracted from multiple hBN crystals on a silicon substrate. A polar material like hBN is generally well-described by a Lorentzian oscillator model of the 'TO-LO' form; therefore, the permittivity of both the in- and out-of-plane optical phonon modes follows:
\begin{equation}\label{eq:epsilon-TO-LO}
  \epsilon_{\mathrm{hBN},s}(\omega) = \epsilon_{\infty,s}\left( \frac{\omega_{\mathrm{LO},s}^2-\omega^2-{\rm i} \omega \Gamma_{s}}{\omega_{\mathrm{TO},s}^2-\omega^2-{\rm i} \omega \Gamma_{s}} \right),
\end{equation}
where $\omega$ is the frequency, $\epsilon_\infty$ is the  high-frequency permittivity, $\omega_{\mathrm{TO}}$ and $\omega_{\mathrm{LO}}$ are the transverse optical (TO) and longitudinal optical (LO) phonon frequencies, and $\Gamma$ is the phonon damping. These parameters depend on the isotopic composition of hBN~\cite{Giles:2018}, necessitating their precise determination.

Employing Raman spectroscopy (LabRAM HR Evolution from HORIBA using a 514\,nm laser source), the E$_\mathrm{2g}$ Raman-active mode, corresponding to the in-plane $\omega_\mathrm{TO}$, was measured and found to be $1364.3\,\mathrm{cm}^{-1}$. The remaining parameters -- namely, the in-plane $\epsilon_\infty$, $\omega_{\mathrm{LO}}$, and $\Gamma$ -- was found with s-SNOM measurements of multiple hBN flakes of varying thickness on silicon substrates.

The parameters of Eq.~(\ref{eq:epsilon-TO-LO}) are summarised in table~\ref{Table:Lorentz}. 
\begin{table}[H]
\caption{
\textbf{In-plane and out-of-plane parameters for the Lorentzian dielectric function of hBN, Eq.~\eqref{eq:epsilon-TO-LO}.}
\label{Table:Lorentz}}
\centering
\begin{tabular}{lcccc}
\hline \hline
				& $\epsilon_\infty$             	
                & $\omega_{\mathrm{LO}}$ 
                & $\omega_{\mathrm{TO}}$
				& $\Gamma$  			\\ 
        		&            	
                &  $(\mathrm{cm}^{-1})$	&  $(\mathrm{cm}^{-1})$
				&  $(\mathrm{cm}^{-1})$ 			\\ \hline
In-plane		& 5.22		& 1619.8 			& 1364.3 		& 3.5   \\
Out-of-plane 	& 2.25		& 820.2  			& 761    		& 2     \\ \hline \hline
\end{tabular}
\end{table}

\bibliographystyle{apsrev4-2}
\bibliography{references}